# Ultra-low-frequency reflection conversion from compressional to shear waves enabled by periodic inclined slits

Kaifei Feng and Pai Peng*

School of Mathematics and Physics, China University of Geosciences, Wuhan 430074, China

*Corresponding author: paipeng@cug.edu.cn

abstract

A periodically patterned free surface with inclined slits can convert an incident compressional wave into a reflected shear wave with nearly complete efficiency at very low frequency. The system is described by two-dimensional in-plane linear elasticity, and the slits are treated as voids. The conversion is quantified by the ratio between the reflected shear-wave energy and the incident compressional-wave energy, obtained from mode decomposition and energy-flux evaluation below the surface. The results indicate that the inclined geometry introduces strong coupling between normal and tangential motions at the boundary, enabling suppression of the ordinary compressional reflection while redirecting the reflected energy into the shear channel. This simple, geometry-controlled mechanism provides a compact route for low-frequency elastic-wave polarization control.

Elastic waves in solids come in different polarizations, and being able to switch between them on demand is useful in a lot of settings, from nondestructive evaluation and ultrasonic inspection to vibration control and even seismic-style wave manipulation. In many practical cases one would like to take an incoming compressional wave and "turn it into" a shear wave, because shear waves can be more sensitive to cracks, interfaces, or anisotropy, and they also carry energy in a different way. The problem is that, in an isotropic solid, polarization conversion is not something you get for free. At a flat traction-free surface, a normally incident compressional wave mainly reflects back as a compressional wave, and the shear channel is essentially shut off. To get strong conversion, the boundary has to provide a mechanism that mixes normal and tangential motion. A common way to promote conversion is to introduce anisotropy, multilayers, or structured interfaces, including metamaterial-type designs. Many of the low-frequency solutions, however, lean on resonant elements to boost the response. Resonance can be effective, but it often comes with a narrow operating band and stronger sensitivity to damping, fabrication tolerances, and parameter drift. At very low frequency the usual alternative is simply to make the structure large, which defeats the point if you want a compact device. However, what is still missing is a compact, geometry-simple interface that can deliver strong compressional-to-shear conversion at very low frequency without relying on resonance; addressing this gap is exactly the contribution of the present work.

In this work, we show that a traction-free surface patterned by a periodic array of inclined void slits can convert a normally incident compressional wave into a reflected shear wave with nearly complete efficiency: the slit inclination induces strong normal–tangential mixing so that compression excites sliding-type motion, and at the right angle and frequency interference suppresses the usual compressional reflection while boosting the shear channel. We model the system using two-dimensional in-plane linear elasticity and quantify conversion by the reflected shear-wave energy relative to the incident compressional-wave energy, extracted from mode decomposition and energy-flux evaluation below the patterned surface, demonstrating that near-complete conversion is achievable in a deep-subwavelength, very low-frequency regime through geometric tuning of the slit angle.

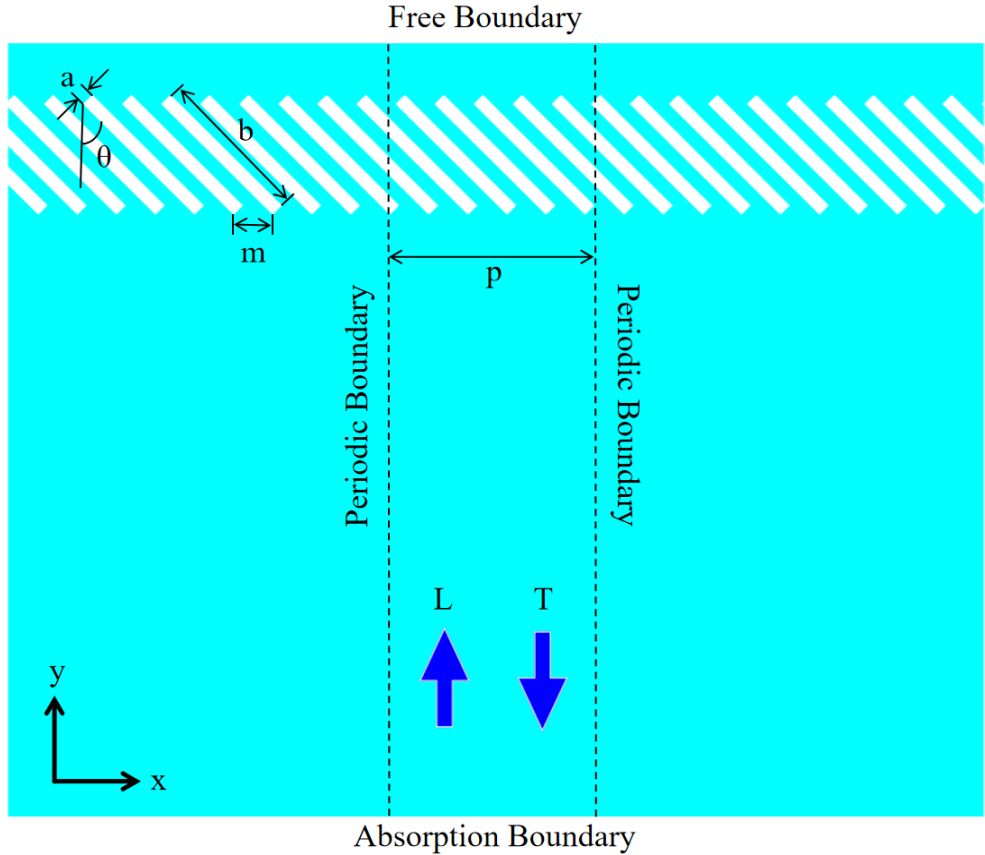

Fig 1. Numerical configuration and geometry of the periodic inclined-slit structure. The slit width a, slit length b, and inclination angle theta (measured from the y axis) define the unit geometry, while m denotes the period and p the lateral size of the simulated domain (p = 10m). A normally incident longitudinal wave impinges from below, and the reflected field is analyzed in terms of longitudinal and transverse components. Boundary labels indicate free, periodic, and absorbing terminations used in the simulations.

We consider a two-dimensional elastodynamic model in the x–y plane with in-plane displacement components (ux, uy). The structure is a periodic array of inclined slits (voids) patterned near the traction-free top surface of an otherwise homogeneous solid. Each slit is characterized by its width a, length b, and inclination angle theta measured with respect to the y axis. The array repeats along the x direction with period m. The computational domain has a lateral width p and contains ten periods, p = 10m. In the representative configuration used throughout this work, we set b = p and theta = 45 degrees; the effective structured height (i.e., the penetration depth of the slit pattern measured in the y direction) is therefore determined jointly by b and theta. The slit region is treated as vacuum (void), so the slit walls are traction-free.

The host medium is modeled as isotropic, linear elastic, and lossless. The material parameters are density 1300 kg/m3, longitudinal wave speed 2540 m/s, and transverse wave speed 1160 m/s. Numerical simulations are carried out in COMSOL Multiphysics. The top boundary is a free boundary (traction-free), including the outer surface and all slit walls. The left and right boundaries are assigned standard periodic boundary conditions to represent an infinitely periodic slit array. The bottom boundary is terminated by a low-reflecting (absorbing) boundary condition to mimic a semi-infinite substrate and suppress spurious reflections from the truncation.

A normally incident longitudinal (P) plane wave is launched from the bottom and propagates upward toward the structured free surface. In a homogeneous sampling region below the slit array, the reflected field is decomposed into a reflected longitudinal component and a reflected transverse (S) component. Practically, this is done by recording the simulated displacement field along a horizontal line in the homogeneous region and separating the two reflected channels using modal projection (or an equivalent plane-wave fitting procedure) so that the reflected transverse-wave energy can be evaluated independently from any residual reflected longitudinal energy.

To quantify mode conversion, we define the conversion ratio, CR, as the energy of the reflected transverse wave divided by the energy of the incident longitudinal wave. With this definition, the maximum possible value is CR = 1, corresponding to complete conversion where all reflected energy returns in the transverse channel. In the next section, we compute the frequency spectrum of CR and focus in particular on the low-frequency regime where ultra-low-frequency conversion is expected.

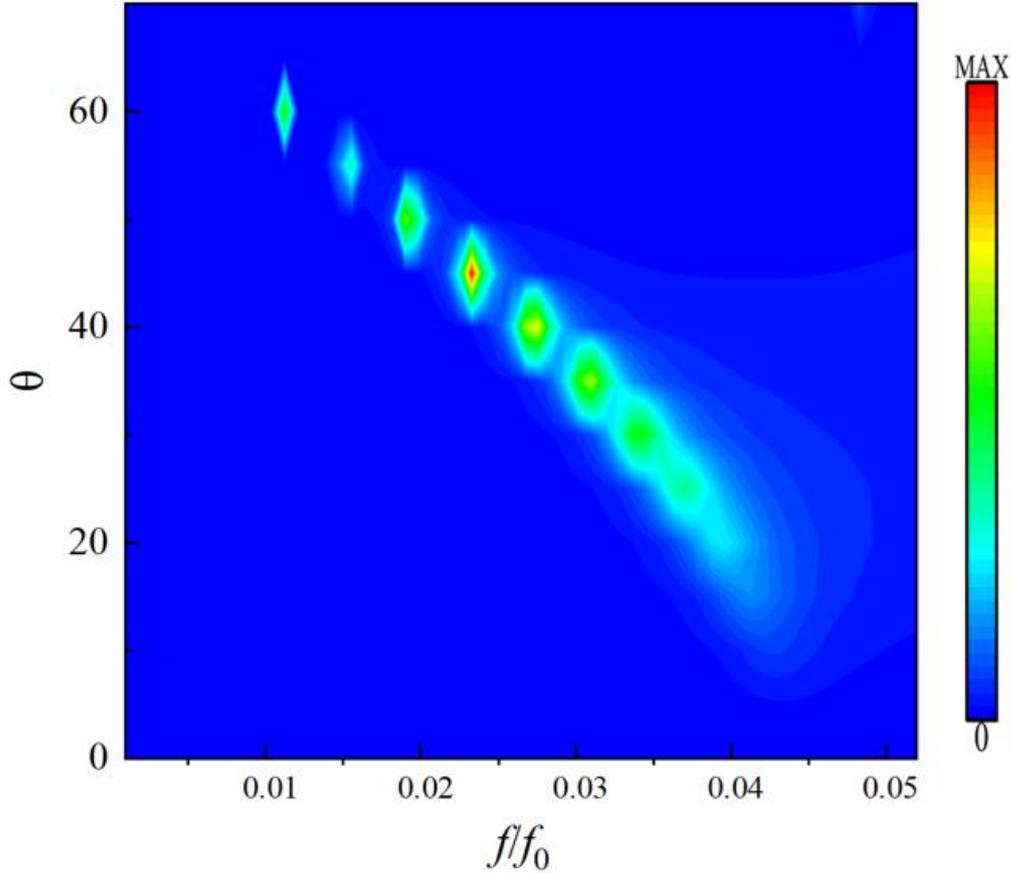

Fig 2. Conversion ratio CR as a function of slit inclination angle theta and normalized frequency f/f0 for the periodic inclined-slit surface with fixed geometry (m = 0.1p, a = 0.01p, b = p). The normalization uses f0 = ct/m, where ct is the shear-wave speed of the host medium. Theta is swept from 0 to 90 degrees in 10-degree steps. CR is defined as the reflected transverse-wave energy divided by the incident longitudinal-wave energy, evaluated from mode decomposition on a horizontal line below the structure and period-averaged over one unit cell.

Figure 2 maps the conversion ratio CR versus theta and the normalized frequency f/f0, with f0 defined as ct/m. Here we keep m =

0.1p, a = 0.01p, and b = p fixed, and only change theta from 0 to 90 degrees (10-degree steps). The main point is that CR can get very close to 1 around f/f0 ≈ 0.02, which is already extremely low. To get a feel for the scale: at f/f0 = 0.02 the shear wavelength is about 50m, and the longitudinal wavelength is larger by the factor cL/ct, so roughly 110m. So the period is tiny compared with either wavelength, and what we see is not a "grating effect" or Bragg thing; it's more like the surface behaves as an effective boundary with some extra degrees of freedom. Why does a tilted-slit surface allow a normally incident P wave to come back mostly as an S wave at all? A flat traction-free surface in an isotropic solid does not really give you a channel to mix the polarizations at normal incidence: you mainly get the usual P reflection. Once you cut the surface with inclined voids, the local deformation is no longer alighned with the global y direction. A vertical compression pushes on the slit walls in a way that naturally produces both opening-type motion and sliding-type motion along the slit direction. That sliding component is basically shear-like, so it feeds the transverse reflected wave. In a simple "effective boundary" picture, the patterned surface is not just softer or stiffer in the normal direction; it also creates cross-coupling between normal and tangential response. That cross-coupling is the ingredient that makes P-to-S conversion possible under normal incidence.

The map also makes it clear that strong conversion is not generic.

Most of the theta–frequency plane stays at low CR, and only a small region gets close to CR = 1 for the present geometry. This is not surprising if you think about what "CR = 1" really demands. It's not only that the structure generates shear; it also needs to kill the ordinary longitudinal reflection at the same time. In practice that means two things have to line up: the longitudinal part of the scattered field must cancel the usual specular P reflection (a destructive interference condition), while the transverse part must radiate efficiently into a propagating S wave rather than staying trapped as near-field deformation around the slits. That second piece is like a matching condition: you want the structure to couple strongly to the S channel and not waste energy in local motion. When both conditions happen to be satisfied together, CR shoots up; when either condition is off, CR stays modest. Theta is the obvious knob because it changes the balance between "mostly normal compliance" and "normal–tangential mixing." For small theta (slits closer to the y axis), the response is closer to vertical voids: you mainly perturb the normal compliance and the tangential component is weak, so conversion is weak. Increasing theta tends to increase the shear-like component driven by a normal load, so the coupling to the reflected S channel improves. But if theta gets too large, the phase relation between the scattered and specular components can drift away from the cancellation condition for the P reflection, or the induced motion may not radiate as cleanly into the far field. That's consistent with seeing only a narrow ridge/hotspot rather than a broad plateau.

One more practical point: theta is sampled very coarsely here (10 degrees). Since the complete-conversion condition is tight, it is quite possible the true optimum sits between two sampled angles, which would make the "almost CR = 1" feature look more isolated than it really is. Still, the central message already stands: near-complete P-to-S conversion can be reached at f/f0 around 0.02 without invoking resonant inclusions, i.e., in a deep-subwavelength, essentially nonresonant setting. Overall, Fig. 2 demonstrates that near-complete P-to-S reflection conversion is achievable in a deep-subwavelength regime, with CR approaching unity already around f/f0 ≈ 0.02 for a suitable theta. This ultra-low-frequency, geometry-controlled conversion is the main result of this work and is summarized in the conclusion below.

In conclusion, we have shown that a simple periodic inclined-slit surface can produce almost complete mode conversion from a normally incident longitudinal (P) wave to a reflected transverse (S) wave. Using a 2D in-plane elastic model for an isotropic, linear elastic host medium and treating the slits as voids, we evaluate the conversion ratio CR defined by the reflected S-wave energy divided by the incident P-wave energy. For the fixed geometry m = 0.1p, a = 0.01p, and b = p, the angle–frequency map reveals a clear ultra-low-frequency working point where CR approaches 1 already around f/f0 ≈ 0.02 (with f0 = ct/m). This confirms that deep-subwavelength, nearly perfect P-to-S reflection conversion is achievable without adding resonant elements. Physically, the inclined slits provide strong normal–tangential coupling at the free surface, which opens the conversion channel and, at the right angle–frequency combination, simultaneously suppresses the ordinary P-wave reflection through interference while allowing efficient radiation into the S-wave channel. The resulting conversion is controlled mainly by geometry (theta) and operates in a very low-frequency regime, which may be useful for low-frequency elastic-wave steering and polarization control in applications where bulky structures are usually required.